\documentclass[11pt]{article}

\usepackage[right=1.0in,left=1.0in,top=1.0in,bottom=1.0in]{geometry}
\usepackage[pagebackref]{hyperref}
\hypersetup{colorlinks, citecolor=blue, filecolor=blue, linkcolor=blue, urlcolor=blue}
\usepackage{graphicx, pgfplots, tikz-network}
\pgfplotsset{compat=1.14}
\pgfplotsset{every axis label/.append style={font=\tiny}}
\usepackage[round]{natbib}
\usepackage{amsmath, amssymb, amsthm}
\usepackage{booktabs, float, xr}
\usepackage[labelsep=period]{caption} 

\newcommand{\m}[1]{\mathbf{#1}}

\newcommand{\s}[1]{\widetilde{#1}}

\newtheorem{theorem}{Theorem}

\newtheorem{assumption}{Assumption}[section]

\newtheorem{definition}{Definition}

\usepackage{setspace}
\usepackage{sectsty}
\sectionfont{\large}
\subsectionfont{\normalsize}
\subsubsectionfont{\normalsize}

\title{Estimating Peer Influence in Multilayer Networks
}

\author{ \\
Weihua An\thanks{Emory University. \href{mailto:weihua.an@emory.edu}{weihua.an@emory.edu}} \and \\
Pablo Estrada\thanks{Capital One. \href{mailto:pabloestradace@gmail.com}{pabloestradace@gmail.com}} \and \\
Juan Estrada\thanks{Analysis Group Economic Consulting. \href{mailto:juan.estrada@analysisgroup.com}{juan.estrada@analysisgroup.com}} \and \\
David Jacho-Chavez\thanks{Emory University. \href{mailto:djachocha@emory.edu}{djachocha@emory.edu}}
}

\date{ \vspace*{0.5cm} \today}   

\hypersetup{
    pdftitle={Estimating Peer Influence in Multilayer Networks},
    pdfauthor={Pablo Estrada},
}

\begin{document}

\bgroup
\let\footnoterule\relax

\begin{singlespace}
\maketitle

\begin{abstract}
    \noindent This paper proposes a model-based empirical method to identify influential individuals in risky behaviors. To determine the most influential individuals, we estimate peer influence using observational cross-sectional data from multiple social connections. Our empirical strategy employs the observed characteristics of distant individuals across multiple social networks as instruments to address the endogeneity arising from homophily. Using Add Health data, we find positive peer effects from friends and classmates on both cigarette smoking and marijuana use. Based on the estimated peer effects, we characterize the influencers in our sample. \\
    \vspace{0in} 
    \vspace{0in}
\end{abstract}
\end{singlespace}
\thispagestyle{empty}
\clearpage

\egroup
\setcounter{page}{1}
\pagebreak

\section{Introduction}\label{sec:intro}

Risky behaviors such as smoking marijuana and tobacco use can adversely affect the well-being and cognitive development of teenagers, according to the U.S. Center for Disease Control.\footnote{\href{https://www.cdc.gov/parents/teens/risk_behaviors.html}{CDC - Teens Risky Behaviors}} Previous literature has established that such behaviors can impact education, crime, future earnings, and healthcare costs.\footnote{For a review of the economic consequences of risky behaviors, see \cite{Cawley_2011_HHE}.} Since peer influence plays a critical role in adolescents' risky behaviors, designing effective prevention programs that account for social connections remains a pressing concern.



Consider a health practitioner aiming to implement a prevention program to reduce risky behaviors among adolescents. How can the practitioner identify the specific adolescents who are the most influential? We assume that conformity is the mechanism through which social interactions influence risky behaviors. Under this assumption, an adolescent's risky behavior tends to align with the average of their connections. However, different types of social links can create varying levels of social influence. For example, adolescents may be more influenced by their friends than their neighbors when deciding whether or not to consume marijuana. Therefore, it may be relevant to incorporate a broader range of possible social connections. In this paper, we demonstrate how, under peer influence and the potentially heterogeneous effects of different types of connections, practitioners can use observational social network data to identify highly influencing individuals with respect to risky behaviors.

We assume that social conformity is the underlying mechanism through which social interactions affect adolescents' risky behaviors. \cite{Ushchev_2020_JET} provide a microfoundation for ``social norms.'' They show that group-based policies are more effective when social norms are the driving force behind individual behavior. Building on their work, we incorporate different types of social connections into the social interactions game. With this social interactions game, we are able to identify highly influential individuals through the development of a novel network centrality measure. Our centrality measure, called the social marginal effects, is related to a heterogeneous Katz-Bonacich centrality, which builds upon the eigenvector centrality for multilayer networks. It incorporates a discount factor for each type of social connection, which is tied to the peer effects specific to that particular connection.

Numerous studies have explored the influence of peers on risky health behaviors, such as marijuana use and smoking. For instance, \cite{Card_2013_RESTAT} estimate a structural model that accounts for peer effects and find that peer influence has a significant impact on marijuana use among adolescents. Similarly, \cite{Arduini_2019_NBER} use the National Longitudinal Study of Adolescent Health to estimate a dynamic social interactions model that considers addiction effects in smoking. These works, along with others,\footnote{See also \cite{Manski_1993_RESTUD}, \cite{Bramoulle_2009_JoE}, \cite{Patacchini_2009_JLEO}, \cite{CalvoArmengol_2009_RESTUD}, \cite{Lin_2010_JLE}, \cite{Liu_2014_JEBO}, \cite{Boucher_2017_EJ}, \cite{Hsieh_2020_JAE}, \cite{Mele_2021_JBES}.} have established the existence of peer effects on risky behaviors. This article contributes to the literature on peer effects by incorporating more than one type of social connections that impact individuals' behavior.

Estimating peer effects using multiple networks poses several methodological challenges, including potential endogeneity issues arising from social networks. Some existing studies \citep[e.g.,][]{GoldsmithPinkham_2013_JBES, Kuersteiner_2020_ECMA} have tackled these challenges using Bayesian and generalized method of moments (GMM) frameworks. Others \citep[e.g.,][]{Comola_2021_RESTAT} have focused on identifying treatment effects when the structure of the network changes. Instead, a related literature studies spillovers assuming that the exposures are exogenous or that the shocks are \citep{Borusyak_2022_RESTUD, Goldsmith_2020_AER}. In our context, exposures are endogenous, since students choose their friendships based on their observed and unobserved characteristics.

This paper presents a new strategy to identify heterogeneous peer effects using multilayer networks, and shows how to recover individuals' social marginal effects with cross-sectional data on multiple social connections. These estimated social effects are then used to identify the influential group of individuals. The main econometric challenge is the endogeneity of the networks that arise from homophily. To address this, we adopt the approach proposed by \cite{Estrada_2021_WP} and utilize the distance of individuals across different networks to construct instruments for identifying and estimating peer effects in a multilayered linear-in-means model \citep{Manta_2022_SPL}. We construct instruments using shocks based on exposures across multiple networks. In our example, we rely on the parental smoking behavior of inter-grade friendships. The resulting estimates are non-linear functions of social effect parameters in a linear model, estimated by a non-linear GMM.

Using data from Add Health, we assess the influence of peers on cigarette smoking and marijuana use, considering two types of social connections: friends and classmates. To avoid ambiguity, what we refer to as the `classmates' network throughout this paper specifically designates a grade-cohort network—that is, students enrolled in the same grade level within their current high school at the time of the survey. Our results reveal positive effects of both friends and classmates on smoking and marijuana use, with the former being more dominant for smoking cigarettes. Notably, the friends effect is twice as large for marijuana use compared to cigarette smoking. Based on our theoretical model, we identify a group of highly influential individuals, who differ from their low-influence counterparts only in terms of their ability test score, gender, and race. 

The contribution of this paper is twofold. First, we propose a novel empirical strategy to identify heterogeneous peer effects using observational cross-sectional data. While previous literature has successfully identified peer influence using natural experiments or longitudinal Stochastic Actor-Oriented Models, these approaches often require experimental conditions or panel network data \citep{Li_et_al_2016, Ivaniushina_et_al_2021, Sijtsema_et_al_2018}. Our approach leverages the multilayer network structure to construct instruments from distant peers, allowing for identification in the more common scenario where only a network snapshot is available. Second, we demonstrate how to use these estimates to calculate social marginal effects for specific individuals. Unlike average treatment effects, this measure provides a granular metric of social influence that identifies key influencers within the network.

The remainder of this article is organized as follows. Section \ref{sec:model} presents the theoretical framework for a model of heterogeneous social interactions. In Section \ref{sec:data}, we describe the data and present descriptive statistics. Section \ref{sec:empirical} discusses the identification strategy and the non-linear GMM estimation. In Section \ref{sec:results}, we present the empirical results and discussion of results. Finally, Section \ref{sec:conclusion} concludes the paper and provides suggestions for future research.

\section{Social Marginal Effects}\label{sec:model}

This section presents a theoretical support for our influence metric: the social marginal effects. To do so, we develop a social interaction model that incorporates social conformity as a determinant of individual preferences. Our model disaggregates the decision to engage in risky behaviors into two components: individual characteristics and social effects. We show that the matrix of social effects yields an index of influence over the risky behavior of others. The social marginal effects can be effectively quantified through a novel measure of heterogeneous Katz-Bonacich centrality that we develop.

\subsection{Social Interactions Model}

Building on the work of \cite{Ushchev_2020_JET} and \cite{Boucher_2024_ECMA}, our model of social interactions captures the influence of social conformity, while also acknowledging the heterogeneity of social connections. To account for the endogeneity of social networks, we employ a two-stage Bayesian game, following the approach of \cite{Blume_2015_JPE}. In the first stage, agents make decisions about their social connections for each type of interaction. Subsequently, they determine the optimal level of risky behavior to engage in. Appendix \ref{app:model} provides a detailed exposition on the social interactions game. The optimal choice of risky behavior for an individual is given by 
\begin{equation}\label{eq:structural}
    \m{y} = \frac{1}{1 + \beta_f^{*} + \beta_c^{*}} \left[ \beta_f^{*} \m{W}_{f} \m{y} + \beta_c^{*} \m{W}_{c} \m{y} + \m{X} \gamma^{*} + \m{e} \right]\text{,}
\end{equation}
which includes social effects stemming from the average outcomes $\m{W}_{f} \m{y}$ and $\m{W}_{c} \m{y}$ of her friends and classmates, respectively. The outcome variable $\m{y}$ is a $n \times 1$ vector, $\m{X}$ is a $n \times k$ matrix that can contain the average characteristics of her connections and individual ones, and $\m{e}$ captures the error term.

In the proposed model, we account for the impact of diverse social connections stemming from $M$ networks, each represented by a row-normalized $(n \times n)$ adjacency matrix $\m{W}_{m}$. The parameters $\beta_m$ quantify an individual's conformity with the average behavior of her peers of type $m$. For instance, $\beta_m$ denotes the social effect of the average cigarette use among friends and classmates when $m=f,c$. Equation \eqref{eq:structural} can be reformulated with the reduced form $\m{y} = \m{S}(\beta^{*}) \left[ \m{X} \gamma^{*} + \m{e} \right]$, where
\begin{equation*}
    \m{S}(\beta^{*}) = \left[\left(1 + \beta^{*}_f + \beta^{*}_c\right)\m{I} - \beta^{*}_f \m{W}_{f} - \beta^{*}_c \m{W}_{c}\right]^{-1}
\end{equation*}
represents a $n \times n$ matrix capturing social effects. The matrix $\m{S}(\beta^{*})$ captures the marginal effect of a change in $j$'s characteristics on $i$'s outcome, as reflected in the entry $s_{ij}$. 

From the structural equation \eqref{eq:structural}, we can derive the regression equation \eqref{eq:regression} that we are seeking to estimate. Equation \eqref{eq:regression} considers solely the two types of social interactions---friends and classmates---that we employ in the empirical analysis. Specifically, we utilize the vector $\m{W}_f \m{y}$ to represent the average cigarette consumption among friends and the vector $\m{W}_c \m{y}$ to denote the average cigarette consumption among classmates. 
In Section \ref{sec:empirical}, we discuss the underlying assumptions that enable the identification of the coefficients $\theta = [\beta_f, \beta_c, \gamma]^{\top}$.
\begin{equation}\label{eq:regression}
    \m{y} = \beta_f \m{W}_f \m{y} + \beta_c \m{W}_c \m{y} + \m{X} \gamma + \m{e}
\end{equation}

The coefficients $\beta_f$ and $\beta_c$ capture the influence of the risky behavior of friends and classmates on their own individual behavior. However, it is important to note that these coefficients should not be conflated with the parameters $\beta^{*}_c$ and $\beta^{*}_f$, which represent the peer effects in the proposed model. Appendix \ref{app:details} describes how to recover the model parameters under Assumption \ref{ass:invertibility} (Invertibility), which restrict the conformity parameters to ensure the existence of the matrix of social effects $\m{S}(\beta^{*})$.

\subsection{Heterogeneous Katz-Bonacich Centrality}

Network centrality is a fundamental concept to determine key agents for influence, dissemination, and strategic actions \citep{Bloch_2023}. In this paper, we define the centrality measure, the social marginal effect of the individual $j$, calculated as $s_j = \sum_{j \neq i} s_{ij}$, where $s_{ij}$ is the $i, j$th element of the matrix $\m{S}(\beta^{*})$. Furthermore, we can compare $s_j$ with the multilayer version of the Katz-Bonacich centrality \citep{Katz_1953, Bonacich_1987}. Among the extensive choices of centrality measures, Katz-Bonacich centrality has been proven effective in contexts of education and politics \citep{CalvoArmengol_2009_RESTUD, Battaglini_2018_JPE}. We first characterize the relationship between the Katz-Bonacich centrality and the monolayer counterpart of our definition of social marginal effects $s_j$. Later in this section, we provide a full description of the heterogeneous Katz-Bonacich centrality.

We first establish the connection between the social marginal effects and the Katz-Bonacich centrality for the case of a single network. \cite{CalvoArmengol_2009_RESTUD} shows that the Katz-Bonacich centrality can be written as $\m{b} = (\m{I} - \beta \m{W})^{-1} (\beta \m{W} \cdot \m{1})$, where $\m{1}$ is a vector of ones and the influence parameter $\beta$ is the non-negative discount factor. The Katz-Bonacich centrality can be expanded as the infinite sum $\m{b} = \sum_{r=1}^{\infty} \beta^r \m{W}^r \cdot \m{1}$. Thus, we define the \textit{indirect} Katz-Bonacich centrality of individual $j$ as
\begin{equation*}
    \s{b}_j = b_j - \sum_{r=1}^{\infty} \beta^r \left[\m{W}^r\right]_{jj} = \sum_{r=1}^{\infty} \beta^r \sum_{i \neq j} \left[\m{W}^r\right]_{ij} ,
\end{equation*}
where $\s{b}_j$ substracts the self-influence and reinforcement effects from the Katz-Bonacich centrality leaving only the inmediate and higher-order effects of others in the network. 

The matrix of marginal effects with a single network $\m{W}$ can be defined as $\m{S}(\beta) = [(1 + \beta)\m{I} - \beta \m{W}]^{-1}$. The social marginal effect $s_j$ is expressed as $\frac{1}{1 + \beta} \sum_{r=0}^{\infty} \left(\frac{\beta}{1 + \beta}\right)^{r} \sum_{i \neq j} \left[\m{W}^r\right]_{ij}$ in the same way. Let the $(1+\beta)-$scaled social marginal effect $\s{s}_j$ be
\begin{equation*}
    \s{s}_j = (1+\beta) s_j = \sum_{r=1}^{\infty} \left( \frac{\beta}{1 + \beta} \right)^{r} \sum_{i \neq j} \left[\m{W}^r\right]_{ij} .
\end{equation*}
Therefore, the scaled social marginal effects $\s{s}_j$ retains the discount factor $\frac{\beta}{1+\beta}$ as part of the weighting. This reflects how distinct social marginal effects decay compared to the indirect Katz-Bonacich centrality $\s{b}_j$.

Previous work by \cite{Sola_2013} showed that the eigenvector centrality can be extended to include multilayer networks. Based on this definition, we relate the social marginal effects to a new measure of Katz-Bonacich centrality over multiple networks. Let $\s{\m{W}} \equiv \sum_{m=1}^M a_m \m{W}_m$ be the multilayer adjacency matrix with weights $a_m$ such that $0 < \sum_{m=1}^M a_m < 1$. Then, the indirect \textit{heterogeneous} Katz-Bonacich centrality and the scaled social marginal effects are
\begin{equation*}
   \s{b}_j = \sum_{r=1}^{\infty} \sum_{i \neq j} \left[\s{\m{W}}^r\right]_{ij} \quad \text{ and } \quad \s{s}_j = \sum_{r=1}^{\infty} \left( \frac{1}{1 + \sum_{m=1}^M a_m} \right)^{r} \sum_{i \neq j} \left[\s{\m{W}}^r\right]_{ij} .
\end{equation*}


The previous expressions highlight that the social marginal effects of our model can be proportional to the heterogeneous Katz-Bonacich centrality. However, instead of having to choose hyperparameters $a_m$ to calculate the heterogeneous Katz-Bonacich centrality as in \cite{Sola_2013}, our social marginal effects rely only on the peer effects $\beta_m$. Therefore, under the assumption that the parameters of the model are identified, this method provides a theoretical justification for the influence metric.

A key advantage of deriving the social marginal effects ($s_j$) from the structural model is its ability to naturally handle multilayer networks. Traditional centrality measures, such as degree or eigenvector centrality, are typically defined for single-layer networks. To apply them in a multilayer context, one would generally have to sum the adjacency matrices (e.g., $W_{total} = W_{friends} + W_{classmates}$), implicitly assuming that a friendship tie carries the exact same influence weight as a classmate tie. However, our empirical results demonstrate that susceptibility varies significantly by layer (i.e., $\beta_f \neq \beta_c$). By incorporating these estimated parameters directly into the centrality calculation, our measure effectively creates a weighted centrality where the weights are determined by the data rather than arbitrary assumptions. This allows us to identify influencers who are not merely `connected,' but who possess ties in the specific layers where peer influence is most powerful.

\section{Data and Descriptive Statistics}\label{sec:data}


\subsection{Add Health Data}

The empirical analysis is based on data from the National Longitudinal Study of Adolescent Health (Add Health), which is a comprehensive, nationally representative survey that gathered data from more than 20,000 adolescents through in-home and in-school interviews \citep{harris_addhealth_2018}. The sample comprises adolescents in grades 7-12 during 1994-95, and the study has tracked them through five waves to date, the latest of which was conducted in 2016-18. We restricted our attention to the saturated subsample of 16 schools where all students were eligible for the in-home questionnaire. Within this subset, we focus exclusively on high school students and their social connections with both friends and classmates.

We examine the impact of peers through two distinct social connections, which we detail in Table \ref{tab:network_stats} in the Appendix \ref{app:tab_fig}. The first is the friendship network, constructed from the Add Health in-home questionnaire, where respondents were asked to nominate up to five female and male friends. We identify reciprocal nominations and construct the corresponding network, which includes 454 pairs and has a density of 0.0005. Despite the low density, the network has an average shortest path of 10 links, providing sufficient variability in connections to investigate peer influence. The construction of the classmates network is based on the in-school questionnaire. We define a directed link from student $i$ to student $j$ if they are same-grade peers. Consequently, this network is composed of multiple disjoint components that are highly clustered.


The two outcomes of interest are cigarette and marijuana use. To measure these outcomes, we use the answers provided in the questionnaires to the following question: ``During the past 30 days, how many days did you smoke \textit{cigarettes/marijuana}?'' To uphold the assumptions of the model, the analysis employs continuous outcome variables to examine the effects of social connections. The individual and parental characteristics of the adolescents are obtained through the in-home questionnaire of the Add Health data. Specifically, we extracted information related to gender, race, age, PVT test (ability) score, physical activity index, parents' characteristics, and the availability of prevention programs in the school from the first wave. In addition, we utilize questions from the second wave that measure risk aversion and future orientation, since these factors are closely associated with potential risky behavior. The inclusion of future orientation is particularly pertinent, as it captures the likelihood that the adolescent believes that they will die at 21 years old.

The parents' characteristics include a dummy for `Two Parents,' which indicates whether the adolescent lives in a household with two parents reflecting family structure stability. The level of education is included through the dummy variable `College Parents.' And a separate dummy for `Smoker Parent,' which indicates the presence of at least one smoker in the household, is used as a control and to construct the main instrumental variable.

We focus on the saturated sample of adolescents to minimize bias from network measurement error. We merged covariates from wave 1 with behavioral indicators from wave 2, resulting in an initial sample size of 2,024 adolescents. To maximize sample retention, we imputed missing PVT test scores for 85 adolescents using the sample mean, and added a missing indicator in all specifications. We subsequently excluded observations with missing data on key controls -- specifically risk aversion, future orientation, and parental smoking -- resulting in a final sample of 1,998 students.

\subsection{Descriptive Statistics}

Our sample consists of 1,998 high school students, of whom 49\% are female. The race variable is constructed using previous work on mixed-race identity \citep{Udry_2003_AJPH}, and the PVT test score is normalized. Future orientation and risk aversion are measured using indices on a scale of 1-5, while physical activity is measured on a scale of 0-3 to indicate how many times the adolescent exercised in the past week. The parents' characteristics consist of dummies for a parent college degree, smokers in the household, and a two-parent household. We also include fixed effects at school-level.

Table \ref{tab:sample_stats} from Appendix \ref{app:tab_fig} displays detailed descriptive statistics of the sample. On average, students in our sample have smoked cigarettes for four days and marijuana for two days in the past month. There is a high variance in the outcomes, particularly with regard to marijuana consumption. The future orientation index is a scale measure of 1-5, with higher values indicating that the adolescent is less future-oriented. In the saturated sample of schools, 32\% and 28\% have a drug and tobacco prevention program, respectively. The dummies for missing values range from 5\% to 7\%, and we include them when estimating the regression equation \eqref{eq:regression}. Our sample has demographics very similar to \cite{Arduini_2019_NBER}.

\section{Empirical Framework}\label{sec:empirical}

The main goal of the paper is to distinguish individuals that bear high social effects. However, estimating social effects presents significant challenges in identifying the underlying peer effects. To overcome the endogeneity of social networks that arise from homophily, we adopt \citeauthor{Estrada_2021_WP}'s \citeyearpar{Estrada_2021_WP} identification strategy. It uses the observed characteristics of distant individuals across different networks as instruments to identify peer effects. In our context, this approach relies on the assumption that the observed characteristics of my friends' classmates, who are not my classmates, are a good instrument for identifying the peer effect of friends.

To estimate peer effects for friends and classmates, we employ a generalized method of moments (GMM) framework. The resulting estimates are then used to calculate the social effects for each adolescent. The GMM estimator incorporates linear and quadratic instruments based on our main identification strategy. \cite{Kuersteiner_2020_ECMA} provide a detailed discussion of our instrument selection and estimation procedure. We derive the social marginal effects and its respective confidence intervals.

\subsection{Identification Strategy}

The literature on peer effects proposes various methods to address the problem of homophily in social interactions, also known as the problem of correlated effects. An approach\footnote{Alternative approaches leverage exogenous variation from field experiments \citep{Carrell_2013_ECMA, Cai_2018_QJE, Chen_2024_RESTAT}, control function approach \citep{Johnsson_2021_RESTAT}, among many others.} to identify the structural parameters of equation \eqref{eq:structural} is to assume that the observable characteristics of individuals are strictly or contemporaneously exogenous. When there is a panel available, \cite{Kuersteiner_2020_ECMA} suggest using past networks, $\m{W}_{m,t-1}$, as instruments for present networks, $\m{W}_{m,t}$. Instead, we will construct instruments based on cross-grade friendships.


Due to homophily, the observable characteristics of individuals are correlated with the unobserved characteristics of peers (friends and classmates), meaning $\mathbb{E}\left[\m{x}_{j} e_{i}\right] \neq 0$. However, we can use distant individuals in the multilayer network to set up moment conditions that help us identify and estimate the structural parameters in equation \eqref{eq:structural}. \cite{Estrada_2021_WP} characterizes distance on multilayer networks using two dimensions: path lengths and edge-type changes. His primary assumption is that individuals connected through different networks are less likely to be dependent than those connected in the same network. In our case, this indicates that individual $i$ is less likely to be similar to her friend's classmates than to her friend's friends.

\begin{figure}
    \centering
    \begin{tikzpicture}
    
    \Vertex[x=0, y=0, color=white, label=2]{2}
    \Vertex[x=1, y=-1, color=white, label=3]{3}
    \Vertex[x=2, y=0, color=gray!50, label=1]{1}
    \Vertex[x=4, y=0, color=white, label=4]{4}
    \Vertex[x=5, y=-1, color=white, label=5]{5}
    \Vertex[x=6, y=0, color=white, label=6]{6}

    \Edge[style={dashed}](1)(2)
    \Edge[style={dashed}](1)(3)
    \Edge[style={dashed}](2)(3)
    \Edge[color=red, bend=45](1)(3)
    \Edge[color=red, bend=-45](2)(3)
    \Edge[color=red, bend=45](1)(4)
    \Edge[style={dashed}](4)(5)
    \Edge[style={dashed}](4)(6)
    \Edge[style={dashed}](5)(6)
    
    \Text[x=1, y=1]{Grade-Cohort 1}
    \Text[x=5, y=1]{Grade-Cohort 2}
    
\end{tikzpicture}
    \caption{Illustration of the identification strategy. Red edges represent friendships and dashed black edges denote same-grade peers (our `classmates' network). In this example, the observed characteristics of nodes 5 and 6 serve as instruments to identify the peer effects of node 1.}
    \label{fig:identification}
\end{figure}
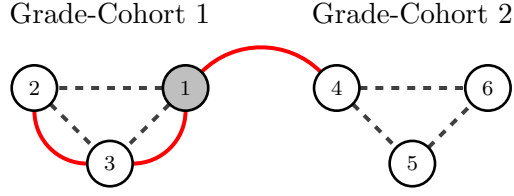

Figure \ref{fig:identification} illustrates how to recover the peer effects stemming from friends ($\beta_f$) on the smoking behavior of individual 1 ($y_1$). Using the observed characteristics of individuals 3 and 4, such as the smoking behavior of their parents ($x_3$ and $x_4$), as instruments for $y_3$ and $y_4$ fails to address the potential endogeneity arising from homophily. Unobserved factors that drive individual 1 to befriend individuals 3 and 4 may also influence their smoking behavior and, by extension, be associated with the smoking behavior of their parents. Furthermore, the exclusion restriction is likely violated: individual 1 may be directly influenced by the smoking behavior of her friends’ parents through interactions in their homes. These two issues---unaccounted homophilic behavior and violation of the exclusion restriction---render such instruments invalid.

Our identification strategy leverages cross-grade friendships using the smoking behavior of the parents of individuals 5 and 6 ($x_3$ and $x_4$) as instruments for $y_4$. If peer effects exist, the smoking behavior of the parents of individuals 5 and 6, which influences the smoking behavior of their children, is correlated with the smoking behavior of 1, only through her friend 4. This can be represented by the orthogonality conditions $\mathbb{E}\left[x_{5} e_{1}\right] = \mathbb{E}\left[x_{6} e_{1}\right] = 0$. Essentially, the nonexistence of a friendship or classmate link between individual 1 and individuals 5 and 6 suppresses concerns on homophilic behavior and omitted variable bias. This strategy holds as long as peer effects exist, which is represented by the relevance condition $\mathbb{E}\left[x_{5} y_{4}\right] \neq 0$ and $\mathbb{E}\left[x_{6} y_{4}\right] \neq 0$.

Our identification strategy circumvents these challenges by leveraging friendships outside of the immediate grade-cohort. Specifically, we use the smoking behavior of the parents of individuals 5 and 6 ($x_5$ and $x_6$) as instruments for $y_4$. If peer effects exist, the smoking behavior of these parents influences the smoking behavior of their children (individuals 5 and 6), which in turn affects individual 4, a direct friend of individual 1. Crucially, there are no direct social or classmate links between individual 1 and individuals 5 and 6, thereby mitigating concerns about homophily and omitted variable bias. This approach satisfies the exclusion restriction, as the smoking behavior of individuals 5 and 6's parents ($x_5$ and $x_6$) affects $y_1$ only through individual 4. The validity of this strategy depends on two key assumptions. First, the orthogonality condition $\mathbb{E}\left[x_{5} e_{1}\right] = \mathbb{E}\left[x_{6} e_{1}\right] = 0$ ensures that the instruments are exogenous. Second, if peer effects exists, the relevance condition $\mathbb{E}\left[x_{5} y_{4}\right] \neq 0$ and $\mathbb{E}\left[x_{6} y_{4}\right] \neq 0$ ensures that the instruments are sufficiently correlated with $y_4$.

Let $\mathcal{W}_f$ be the matrix of instrument assignments with entries $w_{f; i, j} = 1$ if $\mathbb{E}(\m{x}_j e_i) = 0$ is satisfied for agents $i$ and $j$, and $w_{f; i, j} = 0$ otherwise. Under the assumptions in the two-stage Bayesian game of Appendix \ref{app:model}, $\mathbb{E}(\m{x}_j e_i) = 0$ is satisfied when the individual $j$ is a classmate of a friend of the individual $i$. We can follow a similar strategy to construct $\mathcal{W}_c$, but given the assumptions of our model, we can use the same adjacency matrix in the classroom $\m{W}_c$, since it is predetermined. Thus, the matrix of instruments is defined as $\m{Z} = [\mathcal{W}_{f}\widetilde{\m{X}},\  \mathcal{W}_{c}\widetilde{\m{X}},\  \m{X}]$, where $\widetilde{\m{X}}$ are the individual characteristics selected to be instruments such as smoking behavior in the house.

\cite{Estrada_2021_WP} formalizes this identification strategy for the case of $M$ networks. This approach is also related to results on network conductance shown by \cite{Leung_2023_ECMA}. He shows that networks with low conductance (proportion of cluster links sent outside the cluster) satisfy the necessary conditions to conduct cluster-robust inference. In our case, the strength of the instruments depends on the level of network conductance. However, high levels of conductance decrease the likelihood of finding valid instruments.

The identification strategy relies on the assumption that the characteristics of a friend's classmates (who are not the focal student's classmates) affect the focal student only through the friend. A potential limitation is the existence of relationships, such as romantic partnerships, that do not overlap with friendship or classroom ties. For instance, if individual $i$ is in a romantic relationship with individual $k$ (who is a classmate of $i$'s friend $j$), but $i$ does not nominate $k$ as a friend, the exclusion restriction could be violated.

\subsection{Estimation Strategy}

The estimation strategy adopts a GMM framework with linear and quadratic moments. The choice of the functional form of the instruments and the estimation procedure is further explored in \cite{Kuersteiner_2020_ECMA}. To address the endogeneity of social networks, we use the moment restrictions described in the previous subsection to set up matrices $\mathcal{W}_f$ and $\mathcal{W}_c$ as instruments for $\m{W}_f$ and $\m{W}_c$. Thus, the linear and quadratic instruments are
\begin{equation}\label{eq:instruments}
    \m{Z} = \left[ \mathcal{W}_f\widetilde{\m{X}},\  \mathcal{W}_c\widetilde{\m{X}},\  \m{X} \right] \quad \text{ and } \quad \m{A}_m = \mathcal{W}_m^{\top} \mathcal{W}_m - \operatorname{diag} \left( \mathcal{W}_m^{\top} \mathcal{W}_m \right) ,
\end{equation}
where the matrix $\m{Z}$ contains $p$ linear instruments and there are $q$ quadratic instruments by the matrices $\m{A}_m$. For example, $\mathcal{W}_f\widetilde{\m{X}}$ represents the average characteristics of individuals that support the identification strategy, and we use it as an instrument for $\m{W}_f \m{y}$ which is the average risky behavior of her friends. Using these $p+q$ instruments, we write the moment conditions
\begin{equation*}\label{eq:linear_quadratic}
    \overline{\m{m}}_{l}(\theta) = \frac{1}{\sqrt{n}} \left[\m{Z}^{\top} \m{e}(\theta)\right], \qquad \overline{\m{m}}_{q}(\theta) = \frac{1}{\sqrt{n}} \left[\begin{array}{c}
    \m{e}(\theta)^{\top} \m{A}_{f} \m{e}(\theta) \\
    \m{e}(\theta)^{\top} \m{A}_{c} \m{e}(\theta)
    \end{array}\right]\text{.}
\end{equation*}

We concatenate the linear and quadratic instruments as $\overline{\m{m}}_{n}(\theta) = \left[\overline{\m{m}}_{l}(\theta)^{\top},\ \overline{\m{m}}_{q}(\theta)^{\top}\right]^{\top}$, where we stack the regression parameters $\theta = [\beta_f, \beta_c, \gamma]^{\top}$. Thus, the GMM estimator is defined as
\begin{equation*}
    \hat{\theta} = \arg \min _{\theta \in \Theta}\ n^{-1} \overline{\m{m}}_n(\theta)^{\top} \widehat{\m{\Omega}}^{-1}\ \overline{\m{m}}_n(\theta)\text{,}
\end{equation*}
where $\widehat{\m{\Omega}}$ is a moment weighting matrix.

Following \cite{Kuersteiner_2020_ECMA}, we employ the efficient GMM estimator using as a weighting matrix $\widehat{\m{\Omega}} = \left(\m{V}_{l} + 2 \m{V}_{q}\right)$, where $\m{V}_{l} = n^{-1} \m{Z}^{\top} \m{Z}$ and $\m{V}_{q} = n^{-1} \sum_{i=1}^n \sum_{j=1}^n a_{ij}^{\top} a_{ij}$ are extended with zeros to fit the dimensions. Finally, we obtain the variance-covariance matrix $\widehat{\m{\Sigma}} = (\m{G}^{\top} \widehat{\m{\Omega}}^{-1} \m{G})^{-1}$ using the Jacobian matrix $\m{G} = \frac{\partial \overline{\m{m}}_n(\theta)}{\partial \theta}$ of the vector of moments. It is worth noting that this efficient GMM estimator, by optimally weighting the full set of linear and quadratic moment conditions derived from the network structure, often yields estimates with higher precision than standard OLS or 2SLS approaches.


\subsection{Social Effects}

Estimation and inference of social marginal effects require identifying and estimating the parameters $\beta_m$ correctly. The matrix of estimated social marginal effects, denoted as $\m{S}(\hat{\beta})$, is defined as $(1 - \hat{\beta}_f - \hat{\beta}_c)(\m{I} - \hat{\beta}_f \m{W}_f - \hat{\beta}_c \m{W}_c)^{-1}$ when using estimates $\hat{\beta}$. Here, the element $\hat{s}_{ij}$ of this matrix represents $\partial y_i / \partial x_{j,k}$. These effects can amplify or diminish the impacts of any changes in the set of covariates, such as the introduction of a new policy. Therefore, any potential treatment effect would be distorted by the spillovers generated by these social connections.

Our focus is on the aggregate social effects for each individual $j$, which we define as $\hat{s}_j = \sum_{j \neq i} \hat{s}_{ij}$. We interpret this as a measure of the influence that individuals have on their connections. Hence, individuals with high $\hat{s}_j$ are labeled as \emph{influencers}. However, note that this characterization of influence is induced by the peer influence that governs risky behaviors in adolescents. It is worth mentioning that this definition of individual social effects excludes the own marginal effect $s_{ii}$. Since the adjacency matrices $\m{W}_f$ and $\m{W}_c$ are row-normalized, including $s_{ii}$ in the aggregate social effect would amount to $1$ for each individual.

We can also re-scale these aggregate social effects into averages. Since individual social effects depend only on the parameters $\hat{\beta}_f$ and $\hat{\beta}_c$, we can use the delta method to obtain standard errors for the average social effects $\hat{s}_j/n$. The details of the application of the delta method are provided in Appendix \ref{app:details}. To apply the delta method, we rewrite the matrix $\m{S}(\hat{\beta})$ in terms of the infinite sum of its elements. After bounding the expression for $\m{S}(\hat{\beta})$, we apply the delta method to the vector of average social effects $\overline{\m{s}} = n^{-1} [\m{S}(\hat{\beta}) - \operatorname{diag}{\m{S}(\hat{\beta})}]^{\top} \cdot \m{1}$.

\section{Results and Discussion}\label{sec:results}

\subsection{Friends and Classmates Effects}

In this section, we present the empirical findings of our study, which examines the social effects of peers on cigarette smoking and marijuana consumption. Following the identification strategy outlined in Section \ref{sec:empirical}, we utilize the Generalized Method of Moments (GMM) estimation, leveraging cross-grade friendships as instruments based on structural distances across social networks. For comparison, we also estimate the model using the standard Two-Stage Least Squares (2SLS) approach, which assumes the exogeneity of interactions between friends and classmates and relies on networks \(\m{W}^2_f\) and \(\m{W}^2_c\) as instruments \citep{Bramoulle_2009_JoE,Chan_et_al_social_effects,Estrada_Huynh_Jacho_Sanchez_2025,netivreg_stata_journal}. Finally, we include Ordinary Least Squares (OLS) estimates to benchmark against GMM and 2SLS.

A key element of our identification strategy is the use of parental smoking behavior as an instrument of peer influences. This choice allows us to disentangle the effects of friends' and classmates' behaviors on individual smoking and marijuana consumption. We control for a rich set of individual and parental characteristics. Specifically, individual covariates, such as PVT test scores and indicators of high physical activity, high risk aversion, or low future orientation, are allowed to influence the outcomes through both friends and classmates. In contrast, characteristics such as gender, race, age, parental education, school fixed effects, and participation in prevention programs are included in the covariate matrix $\m{X}$ to capture direct effects via the parameter $\gamma$.

Table \ref{tab:cig_effects} displays our main findings on the effects of friends and classmates' peers when analyzing cigarette smoking. The complete set of results, including controls, can be found in Table \ref{tab:cig_effects_all}. Our results reveal significant and precise peer effects for both friends and classmates. Specifically, a one-day increase in the average smoking behavior of an individual's friends is associated with a 0.47 increase in her own smoking behavior. Furthermore, we find substantial effects for classmates that are consistent with other estimates of this type of social interaction. Even without including addiction effects as in \cite{Arduini_2019_NBER}, we obtain very similar classmates' effects, with estimates around 0.35, using our new set of instruments. Notably, the size of our classmates' effects doubles when using our instruments compared to when endogenous networks are employed, increasing from 0.19 to 0.35.

Table \ref{tab:cig_effects} summarizes the estimated peer effects on cigarette smoking. Our analysis reveals significant and precise peer effects for both friends and classmates. Using the GMM estimator, we find that a one-day increase in the average smoking behavior of an individual's friends is associated with a 0.47 increase in her own smoking behavior. This effect is both statistically significant and robust across specifications. Similarly, the estimated influence of classmates on smoking behavior is substantial, with a peer effect of approximately 0.35. Importantly, both peer effects are markedly higher than effects estimated using endogenous networks as instruments. These findings highlight the importance of incorporating exogenous network structures through our identification strategy.

\begin{table}\centering
\setlength{\tabcolsep}{9.5pt}
\def\sym#1{\ifmmode^{#1}\else\(^{#1}\)\fi}
\caption{Peer Effects on Cigarette Smoking}
\begin{tabular}{l|ll|ll|ll}
    \toprule
    {} & \multicolumn{2}{c|}{OLS} & \multicolumn{2}{c|}{2SLS} & \multicolumn{2}{c}{GMM} \\
    \midrule
    \textit{Peer Effects} &  &  &  &  &  &  \\
    Friends Use & 0.396*** & (0.021) & -0.009 & (1.593) & 0.469*** & (0.021) \\
    Classmates Use & 0.178 & (0.151) & 0.191 & (4.162) & 0.351*** & (0.003) \\
    &  &  &  &  &  &  \\
    \textit{Contextual Effects} &  &  &  &  &  &  \\
    Friends Risk & -1.888* & (0.992) & -2.680 & (19.68) & -1.179*** & (0.258) \\
    Classmates Risk & 2.717 & (2.708) & 11.29 & (77.82) & 2.177*** & (0.818) \\
    &  &  &  &  &  &  \\
    \textit{Additional Controls} & \multicolumn{2}{c|}{Yes} & \multicolumn{2}{c|}{Yes} & \multicolumn{2}{c}{Yes} \\
    \textit{Instruments} & \multicolumn{2}{c|}{} & \multicolumn{2}{c|}{Smoker Parent} & \multicolumn{2}{c}{Smoker Parent} \\
    \bottomrule
\end{tabular}

\label{tab:cig_effects}
\vspace{0.25cm}
\begin{minipage}{\textwidth} 
{\footnotesize 
Note: This table reports the peer effect estimates $(\beta_m)$ for Eq. \eqref{eq:regression}. The complete set of results with controls is located in Table \ref{tab:cig_effects_all}. OLS estimation uses clustered standard errors at the school level. 2SLS estimation uses the endogenous networks $\m{W}^2_m$ to calculate peer effects while the GMM estimation incorporates the networks of distant individuals $\mathcal{W}_m$. The GMM estimation uses the efficient variance-covariance matrix for the standard errors.}
\end{minipage}
\end{table}

Our results are consistent with prior research, such as \cite{Arduini_2019_NBER}, which underscores the robustness of our methodology. While we do not explicitly model addiction effects, our estimates align closely with studies that do, suggesting that our approach effectively captures the dynamics of social influence. This reinforces the validity of combining parental smoking behavior and cross-grade friendships as an instrumental variable to identify peer effects.

To further understand the nature of peer effects, we analyze marijuana consumption, as shown in Table \ref{tab:mj_effects}. Here, the influence of friends’ behaviors on marijuana use is reduced by half compared to cigarette smoking, with a peer effect of 0.23. The social influence of classmates also diminishes but remains statistically significant, with an estimated effect of 0.28. These findings align with structural estimates reported in \cite{Card_2013_RESTAT}, which document peer effects ranging from 0.32 to 0.46 for intermediate marijuana use and 0.10 to 0.25 for higher levels of consumption. This consistency serves as a validation of our identification strategy, suggesting that our method successfully recovers structural peer effect parameters using only observational cross-sectional data.

\begin{table}\centering
\setlength{\tabcolsep}{9.5pt}
\def\sym#1{\ifmmode^{#1}\else\(^{#1}\)\fi}
\caption{Peer Effects on Marijuana Use}
\begin{tabular}{l|ll|ll|ll}
    \toprule
    {} & \multicolumn{2}{c|}{OLS} & \multicolumn{2}{c|}{2SLS} & \multicolumn{2}{c}{GMM} \\
    \midrule
    \textit{Peer Effects} &  &  &  &  &  &  \\
    Friends Use & 0.152* & (0.080) & 0.227 & (2.495) & 0.226*** & (0.023) \\
    Classmates Use & -0.067 & (0.231) & 0.005 & (2.421) & 0.281*** & (0.012) \\
    &  &  &  &  &  &  \\
    \textit{Contextual Effects} &  &  &  &  &  &  \\
    Friends Risk & 0.133 & (0.144) & -2.041 & (11.82) & 0.157 & (0.168) \\
    Classmates Risk & -0.786 & (0.795) & 4.241 & (12.98) & 1.172 & (0.976) \\
    &  &  &  &  &  &  \\
    \textit{Additional Controls} & \multicolumn{2}{c|}{Yes} & \multicolumn{2}{c|}{Yes} & \multicolumn{2}{c}{Yes} \\
    \textit{Instruments} & \multicolumn{2}{c|}{} & \multicolumn{2}{c|}{Smoker Parent} & \multicolumn{2}{c}{Smoker Parent} \\
    \bottomrule
\end{tabular}

\label{tab:mj_effects}
\vspace{0.25cm}
\begin{minipage}{\textwidth} 
{\footnotesize 
Note: This table reports the peer effect estimates $(\beta_m)$ for Eq. \eqref{eq:regression}. The complete set of results with controls is located in Table \ref{tab:mj_effects_all}. OLS estimation uses clustered standard errors at the school level. 2SLS estimation uses the endogenous networks $\m{W}^2_m$ to calculate peer effects while the GMM estimation incorporates the networks of distant individuals $\mathcal{W}_m$. The GMM estimation uses the efficient variance-covariance matrix for the standard errors.}
\end{minipage}
\end{table}

An analysis of contextual effects reveals that having friends with a higher risk aversion reduces cigarette smoking, while having no significant effect on marijuana consumption. This is consistent with the notion that individuals with a greater aversion to risk are less likely to engage in behaviors perceived as potentially harmful. Overall, our findings indicate that while the magnitude of peer effects varies across substances, the influence of both friends and classmates remains an important driver of individual behavior.

Tables \ref{tab:cig_effects_all} and \ref{tab:mj_effects_all} in Appendix \ref{app:tab_fig} show the complete set of results for all the controls included. In line with previous research, we find negative effects of smoking among female and black students in the Add Health data. Additionally, we find strong evidence that risk-averse and less future-oriented adolescents engage more in both risky behaviors. Regarding parents' characteristics, having a smoker parent increases cigarette smoking and marijuana use, while having two parents in the household decreases both. However, the effects of having a parent with a college degree differ, with adolescents consuming fewer cigarettes but more marijuana when living with a parent with a college education.

Appendix \ref{app:tab_fig} includes robustness checks for different specifications to assess the sensitivity of our estimates. In Tables \ref{tab:cig_effects_more} and \ref{tab:mj_effects_more}, we include future orientation and physical activity as contextual effects. In addition, we include the same variables as instruments to assess the robustness of the peer effects. Across all specifications, the peer effects remain largely similar. Even when changing school-fixed effects to cohort-fixed effects in Tables \ref{tab:cig_effects_fe} and \ref{tab:mj_effects_fe}, estimates continue to be positive and significant.

\subsection{Identifying Influencers}

Using the peer effects estimates obtained previously, we can now calculate the social effects for the sample of adolescents. Table \ref{tab:influence} presents summary statistics for individual social effects $\hat{s}_j$. The calculated social effects are expected to fall between 0 and 1 since the matrices of social interactions are row-normalized. On average, we find that the social effects of smoking and marijuana consumption are approximately 0.30, which represents the aggregate effect an adolescent has on their peers when there is a change in her own characteristics.

\begin{table}\centering
\def\sym#1{\ifmmode^{#1}\else\(^{#1}\)\fi}
\caption{Adolescents Social Effects on Risky Behaviors}
\begin{tabular}{l|rrrrrrr}
    \toprule
    {} & mean & std & min & 25\% & 50\% & 75\% & max \\
    \midrule
    Cigarette Influence & 0.30 & 0.18 & 0.00 & 0.15 & 0.2 & 0.45 & 0.76 \\
    Marijuana Influence & 0.31 & 0.11 & 0.00 & 0.22 & 0.24 & 0.42 & 0.49 \\
    \bottomrule
\end{tabular}

\label{tab:influence}
\vspace{0.25cm}
\begin{minipage}{0.75\textwidth} 
{\footnotesize 
Note: This table reports summary statistics of the calculated aggregate social effects for each adolescent. We obtain aggregate social effects by summing over the columns of the matrix of social marginal effects $\m{S}(\hat{\beta})$.}
\end{minipage}
\end{table}

There is significant heterogeneity in the aggregate social effect that individuals have. For smoking, the interquartile range of social effects is between 0.15 and 0.45, while for marijuana use, it is between 0.22 and 0.42. Although both behaviors have similar summary statistics for social effects, the upper percentile of the distribution reveals a higher influence on cigarette smoking than on marijuana use. This discrepancy may be due to the stronger effect of friends on smoking cigarettes compared to the use of marijuana.

\begin{figure}\centering
\caption{Histogram of Adolescents Social Effect by Sex}
\scalebox{0.54}{\input{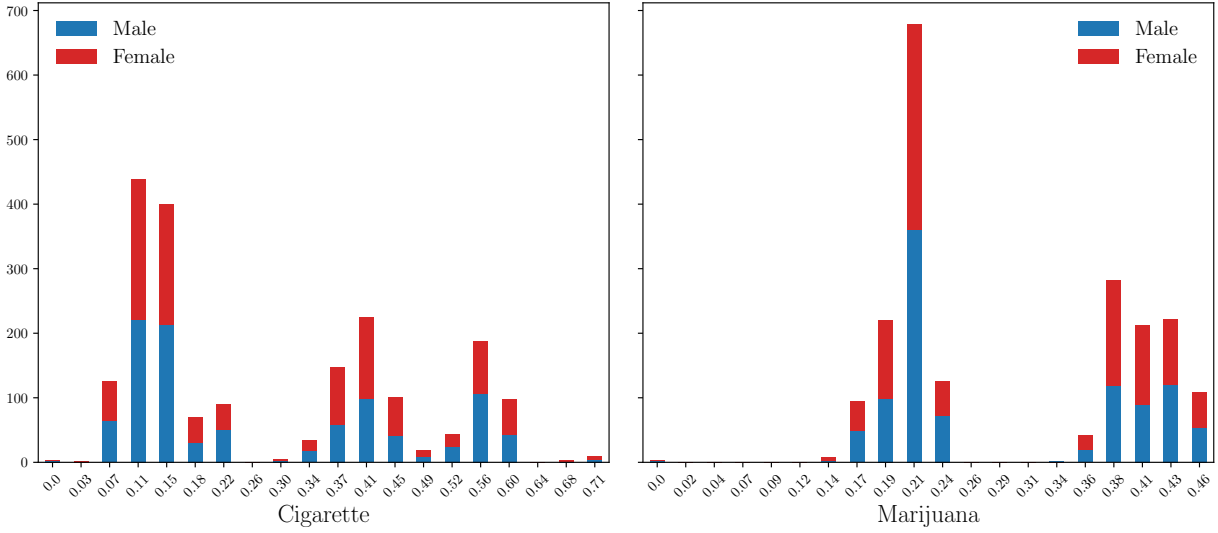}}
\label{fig:influencers_sex}
\vspace{0.25cm}
\begin{minipage}{\textwidth} 
{\footnotesize 
Note: This figure shows the distribution of the aggregate social effects $\hat{s}_j$ for cigarette smoking and marijuana consumption. Each bar of the histogram is divided by the proportion of male and female students.}
\end{minipage}
\end{figure}

Figure \ref{fig:influencers_sex} provides additional insight by displaying the empirical distribution of individual social effects, disaggregated by sex. The histograms show that, for both cigarette smoking and marijuana use, a higher concentration of influencers (those with higher social effects) appears on the right tail of the distribution. Moreover, there seems to be a slightly higher proportion of female influencers compared to males for both risky behaviors. This finding highlights the role of gender in shaping social influence and suggests that policies targeting influencers may need to consider gender-specific interventions.

We classify individuals as `influencers' based on the empirical distribution of their estimated social marginal effects. As shown in Figure \ref{fig:influencers_sex}, the distribution exhibits a heavy right tail, suggesting a natural separation between students with moderate influence and another group with high influence. To rigorously determine this cutoff, we also employed a K-means clustering algorithm on the social effects vector. This unsupervised learning approach identified two distinct clusters separated at the 0.25 cutoff, confirming the separation observed visually. The silhouette score, which measures how well-separated the clusters are, is 0.79 indicating well-clustered points (the closer to 1 the better). Consequently, we define the binary `influencer' variable based on this split, which allows us to clearly compare the demographic and behavioral profiles of these two groups.

Crucially, identifying influencers via social marginal effects does not rely solely on network topology--assuming that the most connected individual is inherently the most influential. In contrast, our measure incorporates the estimated behavioral parameters ($\beta_f, \beta_c$). It accounts for the fact that influence is a function of both connectivity and the susceptibility of an individual's peers. For example, a highly connected student surrounded by peers with low susceptibility would have high centrality but a low social marginal effect. By weighting connections by their estimated conformity, our approach identifies individuals who are not just visible, but who are situated in local networks where social influence is powerful.

We assess the differences in observed characteristics between the two groups of adolescents in Table \ref{tab:differences} for cigarette smoking. There is a statistically significant difference in PVT test scores (IQ test), with the influencers scoring higher than the non-influencers. Cognitive ability may play a role in determining an individual's ability to influence their peers' behaviors. However, the two groups are not significantly different in the amount of physical activity they practice, their future orientation, or their level of risk aversion. The absence of significant differences suggests that influencers and non-influencers share similar behavioral and attitudinal profiles in these dimensions. The lack of variation in risk aversion, in particular, challenges the hypothesis that influencers are more likely to engage in risk-taking behaviors, underscoring the need for more nuanced analyses of the pathways through which social influence manifests. With respect to demographics, the two groups are only significantly different by gender and race. No significant differences were found in their parental characteristics except for living with two parents at home. Similar results were found for marijuana consumption.

Lastly, we examine the relationship between PVT test scores and social effects. Figure \ref{fig:influencers_pvt} displays the social effects for each individual and their test scores. The figure suggests that there is no significant difference between high and low-influence students. However, we observe different slopes for the subgroups in both outcomes. The observed differences in slopes raise important questions about the mechanisms driving social influence. For instance, higher cognitive ability among influencers might amplify their ability to disseminate information, thereby enhancing their impact.  Additionally, using the formula we provide for standard errors on average social effects, we can establish statistically significant differences between groups of individuals.

\subsection{Implications for Adolescent Interventions}

Our findings offer practical guidance for the design of prevention campaigns that rely on peer leaders or similar network-based interventions. Practitioners working with adolescent populations often rely on identifying seed students to disseminate healthy behaviors. Our method leverages the structure of the network and the specific capacity to generate influence among their peers, that is high social marginal effects, to identify and target students.

Specifically, our analysis identifies that high-influence adolescents in this sample tend to have higher cognitive ability scores (PVT). For practitioners, our results suggest that recruiting academic high-performers as peer leaders may be an effective strategy when selecting students to target. Since the driving mechanism in our model is conformity to the average behavior of one's social circle, interventions should focus on changing the public behavior of these high-influence students to shift the local social norm, thereby exerting positive peer influence on the wider student body.

Finally, it is important to acknowledge the temporal context of our data. Our analysis relies on the Add Health saturated sample from the mid-1990s, where the primary form of tobacco consumption was combustible cigarettes. Recent research suggests that modern risky behaviors, such as vaping, may differ from traditional smoking \citep{Valente_et_al_2023}. Therefore, while our methodological contribution, calculating social marginal effects in multilayer networks, is generalizable, the specific behavioral parameters ($\beta$) estimated here should be interpreted within their historical context. Practitioners applying these methods to modern issues like e-cigarette use should re-estimate these parameters using contemporary network data.



\section{Conclusion}\label{sec:conclusion}

This paper investigates how to determine the individuals that have high influence over the network. We study peer influence in the context of high school students and risky behaviors. The empirical strategy uses friends and classmates to create instruments across networks to identify peer effects. We find a group of high- and low-influence individuals, and we show their characteristics by sex and ability test scores.

From a practical standpoint, these results empower practitioners to move beyond broad-spectrum policies and toward targeted network interventions. By using data on friendship and classmate connections, school administrators can apply the social marginal effects algorithm to identify specific influencers within their adolescent populations. Future research should focus on field experiments that compare the efficacy of interventions targeting these structurally identified influencers versus traditional methods of selecting peer leaders.

\section{Acknowledgements}\label{sec:acknowledgement}

This research was previously circulated as ``\emph{Whom to Target under Peer Pressure? A Social Marginal Effects Approach}'' and some results constitute part of Pablo Estrada's doctoral dissertation at Emory University. We thank the editor and two anonymous referees for various helpful comments and suggestions. We are also grateful to Guido Kuersteiner, Paul Goldsmith-Pinkham, and participants in the Network Science and Economics Conference 2023, the Econometric Society European Meeting 2023, and the 2025 Workshop on Causal Network Analysis for their useful feedback. The authors acknowledge financial support from the Emory Program to Enhance Research and Scholarship (PERS). Rachel Clohan provided excellent research assistance.

This research uses data from Add Health (Add Health Restricted-Use Data Contract \#06062201X), a program project directed by Kathleen Mullan Harris and designed by R. Udry, P.S. Bearman, and K.M. Harris at the University of North Carolina at Chapel Hill, and funded by grant 5R01HD047717 from the Eunice Kennedy Shriver National Institute of Child Health and Human Development (NICHD), with cooperative funding from 23 other federal agencies and foundations.

\begin{singlespace}
\def\bibpreamble{The numbers at the end of every reference link to the pages citing the reference.}
\bibliographystyle{aer}   
\bibliography{references}
\end{singlespace}
\clearpage

\appendix

\section{Appendix Tables and Figures}\label{app:tab_fig}
\renewcommand{\tablename}{Appendix Table}
\renewcommand{\thetable}{A\arabic{table}}
\setcounter{table}{0}
\renewcommand{\figurename}{Appendix Figure}
\renewcommand{\thefigure}{A\arabic{figure}}
\setcounter{figure}{0}

\begin{table}[H]\centering
\def\sym#1{\ifmmode^{#1}\else\(^{#1}\)\fi}
\caption{Descriptive Statistics of Networks}
\begin{tabular}{lll}
    \toprule
    Network & Friends & Classmates \\
    \midrule
    Edges & 735 & 182167 \\
    Density & 0.0004 & 0.0913 \\
    Average Cluster & 1.5 & 32.2 \\
    Shortest Path & 12.5 & 1.0 \\
    \bottomrule
\end{tabular}

\label{tab:network_stats}
\end{table}

\vspace{1cm}

\begin{table}[H]\centering
\def\sym#1{\ifmmode^{#1}\else\(^{#1}\)\fi}
\caption{Descriptive Statistics of the Sample}
\begin{tabular}{lll}
    \toprule
    & Mean & SD \\
    \midrule
    Cigarette Use & 4.25 & 9.44 \\
    Friends Cigarette & 1.95 & 6.15 \\
    Classmates Cigarette & 4.25 & 3.42 \\
    Marijuana Use & 1.61 & 7.22 \\
    Friends Marijuana & 0.72 & 4.68 \\
    Classmates Marijuana & 1.61 & 1.29 \\
    Friends Risk & 0.17 & 0.34 \\
    Classmates Risk & 0.41 & 0.10 \\
    Risk Aversion & 0.41 & 0.49 \\
    Future Orientation & 0.49 & 0.50 \\
    Physical Activity & 0.54 & 0.50 \\
    IQ Test & 0.01 & 0.97 \\
    College Parent & 0.65 & 0.48 \\
    Two Parents & 0.72 & 0.45 \\
    Smoker Parent & 0.64 & 0.48 \\
    \bottomrule
\end{tabular}
\label{tab:sample_stats}
\end{table}

\begin{table}[H]\centering
\setlength{\tabcolsep}{9.5pt}
\def\sym#1{\ifmmode^{#1}\else\(^{#1}\)\fi}
\caption{Peer Effects on Cigarette Smoking with Controls}
\begin{tabular}{l|ll|ll|ll}
    \toprule
    {} & \multicolumn{2}{c|}{OLS} & \multicolumn{2}{c|}{2SLS} & \multicolumn{2}{c}{GMM} \\
    \midrule
    \textit{Peer Effects} &  &  &  &  &  &  \\
    Friends Effect & 0.396*** & (0.021) & -0.009 & (1.593) & 0.469*** & (0.021) \\
    Classmates Effect & 0.178 & (0.151) & 0.191 & (4.162) & 0.351*** & (0.003) \\
    &  &  &  &  &  &  \\
    \textit{Contextual Effects} &  &  &  &  &  &  \\
    Friends Risk & -1.888* & (0.992) & -2.680 & (19.68) & -1.179*** & (0.258) \\
    Classmates Risk & 2.717 & (2.708) & 11.29 & (77.82) & 2.177*** & (0.818) \\
    &  &  &  &  &  &  \\
    \textit{Additional Controls}  &  &  &  &  &  &  \\
    Risk Aversion & -1.514*** & (0.488) & -1.523 & (1.247) & -1.684*** & (0.047) \\
    Future Orientation & 0.397* & (0.206) & 0.284 & (0.625) & 0.029 & (0.045) \\
    Physical Activity & -0.472*** & (0.154) & -0.439* & (0.245) & -0.525*** & (0.045) \\
    IQ Test & 0.023 & (0.294) & -0.063 & (1.142) & -0.210*** & (0.028) \\
    Female & 0.156 & (1.086) & 0.229 & (1.830) & 0.377*** & (0.049) \\
    Age & 0.314 & (0.231) & 0.420 & (1.955) & 0.207*** & (0.021) \\
    Black & -1.828*** & (0.326) & -2.077 & (1.952) & 0.136 & (0.089) \\
    Asian & 0.284 & (0.295) & 0.179 & (0.754) & 0.406*** & (0.085) \\
    Other Race & 1.000*** & (0.348) & 0.707 & (1.134) & 3.162*** & (0.119) \\
    College Parent & -0.315 & (0.399) & -0.273 & (0.430) & -1.477*** & (0.051) \\
    Two Parents & -1.062** & (0.418) & -1.342 & (2.115) & -1.747*** & (0.057) \\
    Smoker Parent & 1.827*** & (0.380) & 1.739 & (1.194) & 1.324*** & (0.049) \\
    Fixed Effects & \multicolumn{2}{c|}{School} & \multicolumn{2}{c|}{School} & \multicolumn{2}{c}{School} \\
    &  &  &  &  &  &  \\
    \textit{Instruments} & \multicolumn{2}{c|}{} & \multicolumn{2}{c|}{Smoker Parent} & \multicolumn{2}{c}{Smoker Parent} \\
    \bottomrule
\end{tabular}

\label{tab:cig_effects_all}
\vspace{0.25cm}
\begin{minipage}{\textwidth} 
{\footnotesize 
Note: This table reports the peer effect estimates $(\beta_m)$ for Eq. \eqref{eq:regression}. Each estimation also includes contextual effects for risk aversion; and covariates for asian, other race, drug/tobacco prevention program, dummies for missing PVT test and prevention program, and a constant. OLS estimation uses clustered standard errors at the school level. 2SLS estimation uses the endogenous networks $\m{W}^2_m$ to calculate peer effects while the GMM estimation incorporates the networks of distant individuals $\mathcal{W}_m$. The GMM estimation uses the efficient variance-covariance matrix for the standard errors.}
\end{minipage}
\end{table}

\begin{table}[H]\centering
\setlength{\tabcolsep}{9.5pt}
\def\sym#1{\ifmmode^{#1}\else\(^{#1}\)\fi}
\caption{Peer Effects on Marijuana Use with Controls}
\begin{tabular}{l|ll|ll|ll}
    \toprule
    {} & \multicolumn{2}{c|}{OLS} & \multicolumn{2}{c|}{2SLS} & \multicolumn{2}{c}{GMM} \\
    \midrule
    \textit{Peer Effects} &  &  &  &  &  &  \\
    Friends Effect & 0.152* & (0.080) & 0.227 & (2.495) & 0.226*** & (0.023) \\
    Classmates Effect & -0.067 & (0.231) & 0.005 & (2.421) & 0.281*** & (0.012) \\
    &  &  &  &  &  &  \\
    \textit{Contextual Effects} &  &  &  &  &  &  \\
    Friends Risk & 0.133 & (0.144) & -2.041 & (11.82) & 0.157 & (0.168) \\
    Classmates Risk & -0.786 & (0.795) & 4.241 & (12.98) & 1.172 & (0.976) \\
    &  &  &  &  &  &  \\
    \textit{Additional Controls}  &  &  &  &  &  &  \\
    Risk Aversion & -1.059*** & (0.262) & -1.068*** & (0.368) & -1.132*** & (0.047) \\
    Future Orientation & 1.000*** & (0.227) & 0.974** & (0.469) & 0.791*** & (0.045) \\
    Physical Activity & -0.025 & (0.499) & -0.017 & (0.467) & 0.352*** & (0.045) \\
    IQ Test & -0.100 & (0.102) & -0.061 & (0.083) & 0.259*** & (0.026) \\
    Female & -1.209*** & (0.239) & -1.052 & (1.376) & -0.973*** & (0.048) \\
    Age & 0.012 & (0.159) & -0.009 & (0.187) & -0.059*** & (0.022) \\
    Black & -0.763*** & (0.113) & -0.739* & (0.415) & 0.253*** & (0.086) \\
    Asian & -1.913*** & (0.213) & -1.801 & (2.163) & -1.903*** & (0.087) \\
    Other Race & -0.301 & (0.246) & -0.347 & (1.387) & 1.001*** & (0.122) \\
    College Parent & 0.372 & (0.268) & 0.410 & (0.816) & 0.761*** & (0.052) \\
    Two Parents & -0.563** & (0.265) & -0.458 & (1.113) & -0.730*** & (0.054) \\
    Smoker Parent & 0.595*** & (0.184) & 0.473 & (0.818) & 0.409*** & (0.049) \\
    Fixed Effects & \multicolumn{2}{c|}{School} & \multicolumn{2}{c|}{School} & \multicolumn{2}{c}{School} \\
    &  &  &  &  &  &  \\
    \textit{Instruments} & \multicolumn{2}{c|}{} & \multicolumn{2}{c|}{Smoker Parent} & \multicolumn{2}{c}{Smoker Parent} \\
    \bottomrule
\end{tabular}

\label{tab:mj_effects_all}
\vspace{0.25cm}
\begin{minipage}{\textwidth} 
{\footnotesize 
Note: This table reports the peer effect estimates $(\beta_m)$ for Eq. \eqref{eq:regression}. Each estimation also includes contextual effects for risk aversion; and covariates for asian, other race, drug/tobacco prevention program, dummies for missing PVT test and prevention program, and a constant. OLS estimation uses clustered standard errors at the school level. 2SLS estimation uses the endogenous networks $\m{W}^2_m$ to calculate peer effects while the GMM estimation incorporates the networks of distant individuals $\mathcal{W}_m$. The GMM estimation uses the efficient variance-covariance matrix for the standard errors.}
\end{minipage}
\end{table}

\begin{table}[H]\centering
\setlength{\tabcolsep}{9.5pt}
\def\sym#1{\ifmmode^{#1}\else\(^{#1}\)\fi}
\caption{Peer Effects on Cigarette Smoking with other Instruments}
\begin{tabular}{l|ll|ll|ll}
    \toprule
    {} & \multicolumn{2}{c|}{OLS} & \multicolumn{2}{c|}{2SLS} & \multicolumn{2}{c}{GMM} \\
    \midrule
    \textit{Peer Effects} &  &  &  &  &  &  \\
    Friends Effect & 0.420*** & (0.019) & 0.171 & (0.279) & 0.480*** & (0.021) \\
    Classmates Effect & 0.132 & (0.161) & 0.547*** & (0.205) & 0.316*** & (0.008) \\
    &  &  &  &  &  &  \\
    \textit{Contextual Effects} &  &  &  &  &  &  \\
    Friends Risk & -0.777 & (0.957) & -1.279 & (4.686) & -0.510 & (0.689) \\
    Friends Future & -1.935** & (0.965) & -1.819 & (4.632) & -0.511 & (0.363) \\
    Friends Physical & -0.991 & (0.653) & -0.542 & (2.829) & 4.050*** & (0.290) \\
    Classmates Risk & 2.162 & (2.851) & 1.642 & (3.106) & -1.773*** & (0.445) \\
    Classmates Future & 3.283* & (1.889) & 0.598 & (2.166) & 1.639*** & (0.391) \\
    Classmates Physical & -2.385 & (2.024) & -0.696 & (1.787) & -1.801*** & (0.385) \\
    &  &  &  &  &  &  \\
    \textit{Additional Controls}  &  &  &  &  &  &  \\
    Risk Aversion & -1.493*** & (0.500) & -1.598*** & (0.480) & -1.647*** & (0.050) \\
    Future Orientation & 0.425** & (0.199) & 0.307 & (0.280) & 0.067 & (0.047) \\
    Physical Activity & -0.471*** & (0.171) & -0.408* & (0.216) & -0.409*** & (0.048) \\
    IQ Test & 0.037 & (0.311) & -0.015 & (0.367) & -0.157*** & (0.027) \\
    Female & 0.184 & (1.097) & 0.189 & (1.061) & 0.418*** & (0.051) \\
    Age & 0.181 & (0.287) & 0.119 & (0.199) & 0.065*** & (0.023) \\
    Black & -2.062*** & (0.290) & -2.258*** & (0.396) & -0.440*** & (0.097) \\
    Asian & 0.387 & (0.287) & 0.323 & (0.447) & 0.528*** & (0.085) \\
    Other Race & 0.814** & (0.334) & 0.762** & (0.322) & 2.625*** & (0.120) \\
    College Parent & -0.266 & (0.380) & -0.240 & (0.364) & -1.431*** & (0.051) \\
    Two Parents & -1.025** & (0.409) & -1.216** & (0.495) & -1.748*** & (0.056) \\
    Smoker Parent & 1.844*** & (0.366) & 1.805*** & (0.417) & 1.414*** & (0.050) \\
    Fixed Effects & \multicolumn{2}{c|}{School} & \multicolumn{2}{c|}{School} & \multicolumn{2}{c}{School} \\
    &  &  &  &  &  &  \\
    \textit{Instruments} & \multicolumn{2}{c|}{} & \multicolumn{2}{c|}{Smoker Parent} & \multicolumn{2}{c}{Smoker Parent} \\
    & \multicolumn{2}{c|}{} & \multicolumn{2}{c|}{Risk Aversion} & \multicolumn{2}{c}{Risk Aversion} \\
    & \multicolumn{2}{c|}{} & \multicolumn{2}{c|}{Future Orientation} & \multicolumn{2}{c}{Future Orientation} \\
    & \multicolumn{2}{c|}{} & \multicolumn{2}{c|}{Physical Activity} & \multicolumn{2}{c}{Physical Activity} \\
    \bottomrule
\end{tabular}

\label{tab:cig_effects_more}
\vspace{0.25cm}
\begin{minipage}{\textwidth} 
{\footnotesize 
Note: This table reports the peer effect estimates $(\beta_m)$ for Eq. \eqref{eq:regression}. Each estimation also includes contextual effects for physical activity, future orientation, and risk aversion; and covariates for asian, other race, drug/tobacco prevention program, dummies for missing PVT test and prevention program, and a constant. OLS estimation uses clustered standard errors at the school level. 2SLS estimation uses the endogenous networks $\m{W}^2_m$ to calculate peer effects while the GMM estimation incorporates the networks of distant individuals $\mathcal{W}_m$. The GMM estimation uses the efficient variance-covariance matrix for the standard errors.}
\end{minipage}
\end{table}

\begin{table}[H]\centering
\setlength{\tabcolsep}{9.5pt}
\def\sym#1{\ifmmode^{#1}\else\(^{#1}\)\fi}
\caption{Peer Effects on Marijuana Use with other Instruments}
\begin{tabular}{l|ll|ll|ll}
    \toprule
    {} & \multicolumn{2}{c|}{OLS} & \multicolumn{2}{c|}{2SLS} & \multicolumn{2}{c}{GMM} \\
    \midrule
    \textit{Peer Effects} &  &  &  &  &  &  \\
    Friends Effect & 0.163* & (0.084) & 0.639*** & (0.147) & 0.231*** & (0.022) \\
    Classmates Effect & -0.092 & (0.255) & 0.248 & (0.273) & 0.252*** & (0.013) \\
    &  &  &  &  &  &  \\
    \textit{Contextual Effects} &  &  &  &  &  &  \\
    Friends Risk & 0.543** & (0.251) & -0.500 & (4.080) & 0.390 & (0.879) \\
    Friends Future & -0.928** & (0.383) & -1.816 & (1.925) & 0.189 & (0.564) \\
    Friends Physical & -0.100 & (0.385) & -0.136 & (2.935) & 3.275*** & (0.278) \\
    Classmates Risk & -0.999 & (0.881) & 0.501 & (1.706) & -0.929** & (0.422) \\
    Classmates Future & 1.655 & (1.134) & 1.101 & (0.818) & 1.496*** & (0.484) \\
    Classmates Physical & -1.620 & (0.987) & -2.403 & (1.494) & -2.450*** & (0.381) \\
    &  &  &  &  &  &  \\
    \textit{Additional Controls}  &  &  &  &  &  &  \\
    Risk Aversion & -1.059*** & (0.262) & -1.098*** & (0.243) & -1.139*** & (0.052) \\
    Future Orientation & 1.014*** & (0.224) & 1.069*** & (0.227) & 0.857*** & (0.047) \\
    Physical Activity & -0.046 & (0.499) & -0.036 & (0.521) & 0.333*** & (0.050) \\
    IQ Test & -0.105 & (0.099) & -0.106 & (0.126) & 0.265*** & (0.026) \\
    Female & -1.203*** & (0.251) & -0.961*** & (0.280) & -0.976*** & (0.055) \\
    Age & -0.055 & (0.199) & -0.131 & (0.162) & -0.125*** & (0.024) \\
    Black & -0.854*** & (0.116) & -0.900** & (0.415) & 0.068 & (0.105) \\
    Asian & -1.870*** & (0.215) & -1.444*** & (0.083) & -1.817*** & (0.086) \\
    Other Race & -0.353 & (0.217) & 0.005 & (0.136) & 0.968*** & (0.124) \\
    College Parent & 0.396 & (0.284) & 0.569** & (0.240) & 0.767*** & (0.052) \\
    Two Parents & -0.557** & (0.269) & -0.314 & (0.251) & -0.740*** & (0.055) \\
    Smoker Parent & 0.603*** & (0.166) & 0.457** & (0.184) & 0.427*** & (0.053) \\
    Fixed Effects & \multicolumn{2}{c|}{School} & \multicolumn{2}{c|}{School} & \multicolumn{2}{c}{School} \\
    &  &  &  &  &  &  \\
    \textit{Instruments} & \multicolumn{2}{c|}{} & \multicolumn{2}{c|}{Smoker Parent} & \multicolumn{2}{c}{Smoker Parent} \\
    & \multicolumn{2}{c|}{} & \multicolumn{2}{c|}{Risk Aversion} & \multicolumn{2}{c}{Risk Aversion} \\
    & \multicolumn{2}{c|}{} & \multicolumn{2}{c|}{Future Orientation} & \multicolumn{2}{c}{Future Orientation} \\
    & \multicolumn{2}{c|}{} & \multicolumn{2}{c|}{Physical Activity} & \multicolumn{2}{c}{Physical Activity} \\
    \bottomrule
\end{tabular}

\label{tab:mj_effects_more}
\vspace{0.25cm}
\begin{minipage}{\textwidth} 
{\footnotesize 
Note: This table reports the peer effect estimates $(\beta_m)$ for Eq. \eqref{eq:regression}. Each estimation also includes contextual effects for physical activity, future orientation, and risk aversion; and covariates for asian, other race, drug/tobacco prevention program, dummies for missing PVT test and prevention program, and a constant. OLS estimation uses clustered standard errors at the school level. 2SLS estimation uses the endogenous networks $\m{W}^2_m$ to calculate peer effects while the GMM estimation incorporates the networks of distant individuals $\mathcal{W}_m$. The GMM estimation uses the efficient variance-covariance matrix for the standard errors.}
\end{minipage}
\end{table}

\begin{table}[H]\centering
\setlength{\tabcolsep}{9.5pt}
\def\sym#1{\ifmmode^{#1}\else\(^{#1}\)\fi}
\caption{Peer Effects on Cigarette with other Fixed Effects}
\begin{tabular}{l|ll|ll|ll}
    \toprule
    {} & \multicolumn{2}{c|}{No FE} & \multicolumn{2}{c|}{School FE} & \multicolumn{2}{c}{Cohort FE} \\
    \midrule
    \textit{Peer Effects} &  &  &  &  &  &  \\
    Friends Effect & 0.472*** & (0.014) & 0.469*** & (0.021) & 0.474*** & (0.020) \\
    Classmates Effect & 0.419*** & (0.002) & 0.331*** & (0.004) & 0.427*** & (0.007) \\
    &  &  &  &  &  &  \\
    \textit{Contextual Effects} &  &  &  &  &  &  \\
    Friends Risk & -1.145*** & (0.187) & -1.142*** & (0.262) & -1.032*** & (0.241) \\
    Classmates Risk & 2.418*** & (0.311) & 2.215** & (0.870) & 2.521*** & (0.478) \\
    &  &  &  &  &  &  \\
    \textit{Additional Controls}  &  &  &  &  &  &  \\
    Risk Aversion & -1.616*** & (0.046) & -1.711*** & (0.047) & -1.599*** & (0.046) \\
    Future Orientation & 0.068 & (0.045) & -0.002 & (0.045) & 0.005 & (0.045) \\
    Physical Activity & -0.460*** & (0.044) & -0.566*** & (0.045) & -0.442*** & (0.045) \\
    IQ Test & -0.201*** & (0.025) & -0.190*** & (0.028) & -0.342*** & (0.027) \\
    Female & 0.222*** & (0.049) & 0.380*** & (0.049) & -0.138*** & (0.050) \\
    Age & 0.144*** & (0.007) & 0.258*** & (0.022) & -0.872*** & (0.042) \\
    Black & -0.580*** & (0.077) & 0.108 & (0.089) & -0.766*** & (0.077) \\
    Asian & 0.041 & (0.078) & 0.322*** & (0.085) & 0.027 & (0.080) \\
    Other Race & 2.670*** & (0.115) & 3.107*** & (0.119) & 2.278*** & (0.116) \\
    College Parent & -1.541*** & (0.051) & -1.406*** & (0.051) & -1.541*** & (0.051) \\
    Two Parents & -1.904*** & (0.058) & -1.700*** & (0.058) & -1.979*** & (0.057) \\
    Smoker Parent & 1.470*** & (0.048) & 1.309*** & (0.049) & 1.537*** & (0.049) \\
    &  &  &  &  &  &  \\
    \textit{Instruments} & \multicolumn{2}{c|}{Smoker Parent} & \multicolumn{2}{c|}{Smoker Parent} & \multicolumn{2}{c}{Smoker Parent} \\
    \bottomrule
\end{tabular}

\label{tab:cig_effects_fe}
\vspace{0.25cm}
\begin{minipage}{\textwidth} 
{\footnotesize 
Note: This table reports the peer effect estimates $(\beta_m)$ for Eq. \eqref{eq:regression}. Each estimation also includes contextual effects for risk aversion; and covariates for asian, other race, drug/tobacco prevention program, dummies for missing PVT test and prevention program, and a constant. OLS estimation uses clustered standard errors at the school level. 2SLS estimation uses the endogenous networks $\m{W}^2_m$ to calculate peer effects while the GMM estimation incorporates the networks of distant individuals $\mathcal{W}_m$. The GMM estimation uses the efficient variance-covariance matrix for the standard errors.}
\end{minipage}
\end{table}

\begin{table}[H]\centering
\setlength{\tabcolsep}{9.5pt}
\def\sym#1{\ifmmode^{#1}\else\(^{#1}\)\fi}
\caption{Peer Effects on Marijuana Use with other Fixed Effects}
\begin{tabular}{l|ll|ll|ll}
    \toprule
    {} & \multicolumn{2}{c|}{No FE} & \multicolumn{2}{c|}{School FE} & \multicolumn{2}{c}{Cohort FE} \\
    \midrule
    \textit{Peer Effects} &  &  &  &  &  &  \\
    Friends Effect & 0.229*** & (0.029) & 0.227*** & (0.023) & 0.229*** & (0.027) \\
    Classmates Effect & 0.335*** & (0.012) & 0.280*** & (0.012) & 0.321*** & (0.022) \\
    &  &  &  &  &  &  \\
    \textit{Contextual Effects} &  &  &  &  &  &  \\
    Friends Risk & 0.202 & (0.174) & 0.167 & (0.168) & 0.154 & (0.182) \\
    Classmates Risk & 0.448 & (0.291) & 1.139 & (0.983) & -0.754** & (0.341) \\
    &  &  &  &  &  &  \\
    \textit{Additional Controls}  &  &  &  &  &  &  \\
    Risk Aversion & -1.090*** & (0.046) & -1.132*** & (0.048) & -1.169*** & (0.046) \\
    Future Orientation & 0.766*** & (0.045) & 0.792*** & (0.046) & 0.805*** & (0.046) \\
    Physical Activity & 0.441*** & (0.044) & 0.352*** & (0.045) & 0.420*** & (0.045) \\
    IQ Test & 0.167*** & (0.025) & 0.260*** & (0.026) & 0.197*** & (0.026) \\
    Female & -0.942*** & (0.048) & -0.973*** & (0.048) & -0.967*** & (0.050) \\
    Age & 0.075*** & (0.007) & -0.059*** & (0.022) & 0.283*** & (0.040) \\
    Black & 0.426*** & (0.068) & 0.251*** & (0.086) & 0.682*** & (0.072) \\
    Asian & -1.548*** & (0.079) & -1.902*** & (0.087) & -1.425*** & (0.081) \\
    Other Race & 1.588*** & (0.115) & 0.997*** & (0.122) & 1.928*** & (0.118) \\
    College Parent & 0.700*** & (0.053) & 0.763*** & (0.052) & 0.732*** & (0.053) \\
    Two Parents & -0.716*** & (0.055) & -0.730*** & (0.054) & -0.796*** & (0.056) \\
    Smoker Parent & 0.427*** & (0.049) & 0.412*** & (0.049) & 0.362*** & (0.049) \\
    &  &  &  &  &  &  \\
    \textit{Instruments} & \multicolumn{2}{c|}{Smoker Parent} & \multicolumn{2}{c|}{Smoker Parent} & \multicolumn{2}{c}{Smoker Parent} \\
    \bottomrule
\end{tabular}

\label{tab:mj_effects_fe}
\vspace{0.25cm}
\begin{minipage}{\textwidth} 
{\footnotesize 
Note: This table reports the peer effect estimates $(\beta_m)$ for Eq. \eqref{eq:regression}. Each estimation also includes contextual effects for risk aversion; and covariates for asian, other race, drug/tobacco prevention program, dummies for missing PVT test and prevention program, and a constant. OLS estimation uses clustered standard errors at the school level. 2SLS estimation uses the endogenous networks $\m{W}^2_m$ to calculate peer effects while the GMM estimation incorporates the networks of distant individuals $\mathcal{W}_m$. The GMM estimation uses the efficient variance-covariance matrix for the standard errors.}
\end{minipage}
\end{table}

\begin{table}[H]\centering
\def\sym#1{\ifmmode^{#1}\else\(^{#1}\)\fi}
\caption{Difference between High and Low-Influencing Adolescents}
\begin{tabular}{ll}
    \toprule
    & Difference \\
    \midrule
    Risk Aversion & -0.03 \\
    Future Orientation & -0.03 \\
    Physical Activity & -0.00 \\
    IQ Test & \ 0.29*** \\
    Female & \ 0.05** \\
    Age & -0.08 \\
    Black & -0.10*** \\
    Asian & -0.00 \\
    Other Race & -0.04*** \\
    College Parent & \ 0.02 \\
    Two Parents & \ 0.06*** \\
    Smoker Parent & \ 0.03 \\
    \bottomrule
\end{tabular}
\label{tab:differences}
\vspace{0.25cm} \\
\begin{minipage}{0.6\textwidth} 
{\footnotesize 
Note: This table reports differences in means for high-influencing individuals vs low-influencing individuals on cigarette smoking. The p-value is calculated using a t-test for two independent samples with different variances.}
\end{minipage}
\end{table}

\begin{figure}[H]\centering
\caption{Adolescents Social Effect and PVT Test Scores (normalized)}
\scalebox{0.52}{\input{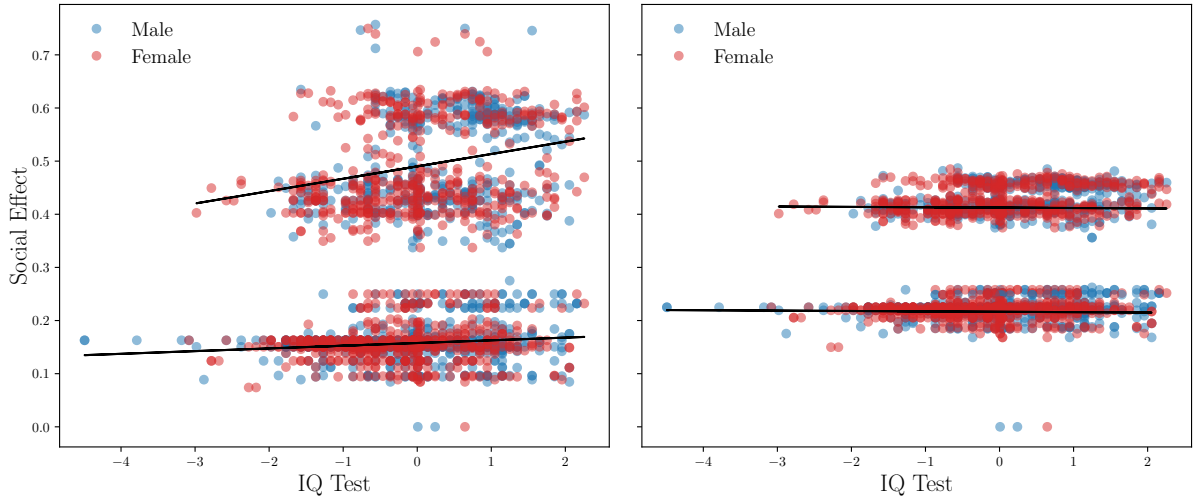}}
\label{fig:influencers_pvt}
\begin{minipage}{\textwidth} 
{\footnotesize 
Note: This figure shows aggregate social effects $\hat{s}_j$ and PVT (IQ test) scores for cigarette smoking in the left and marijuana consumption in the right panel. Each individual dot is colored by gender and the linear relationship is plotted by a black line between subgroups.}
\end{minipage}
\end{figure}



\newpage

\section{Social Interactions Model}\label{app:model}

In this section, we present a model of social interactions in which preferences feature  social conformity with heterogeneous effects depending on the type of social connection. Following \cite{Blume_2015_JPE}, we assume that individuals exhibit a quadratic utility function that highlights contextual effects, peer effects, and the cost of taking action. The setup of the game has two stages. First, there is a network formation process for $\{G_m\}^M_{m=1}$ networks. The networks are characterized by a set of nodes $V$ that remain constant and edges $E_m$ that vary for each layer $m$. In the second stage, agents choose the optimal action of risky behavior $y_i$. We solve the game by backward induction. First, we solve for the optimal action and then for the network formation process given the optimal actions.

\vspace{0.3cm}

\noindent \textbf{Second Stage.} The set $V = \{1, \dots, n\}$ denotes the players in the game with individual characteristics $\left(x_{i}, b_{i}\right)$. Characteristics $x_{i}$, such as physical activity and grades, are publicly observed. Instead, the characteristics $b_{i}$ are private information about each individual that only they know about themselves. An example of this private information could be a family history of substance abuse. Individual types are defined as $\left(x, b_{i}\right) \in \mathbb{R}^{n+1}$. Thus, individual $i$ chooses action $y_{i} \in \mathbb{R}$ to maximize
\begin{equation*}
    \begin{aligned}
    U_{i}\left(y_{i},\ y_{-i}\right) = & \left(\gamma^{*} x_{i}+ b_{i} + \sum_{m=1}^M \sum_{j=1}^n w_{ij,m}\ x_{j}\ \delta^{*}_{m}\right) y_{i} \\
    & - \frac{1}{2} y_{i}^{2} -  \frac{1}{2} \sum_{m=1}^M \beta^{*}_{m} \left( y_{i} - \sum_{j=1}^n w_{ij,m}\ y_{j} \right)^2 , 
\end{aligned}
\end{equation*}
and the best response function\footnote{Details of the proof can be found in \cite{Blume_2015_JPE} and \cite{Estrada_2021_WP} but for the case when preferences display strategic complementarities. The proof dependends on imposing a prior distribution on players' private types, which is exogenous and of common knowledge.} is
\begin{equation*}
    \m{y} = \left[\left(1 + \sum_{m=1}^M \beta^{*}_m\right)\m{I} - \sum_{m=1}^M \beta^{*}_m \m{W}_{m} \right]^{-1} \left[ \sum_{m=1}^M \delta^{*}_m \m{W}_{m} \m{x} + \gamma^{*} \m{x} + \mu(\m{x}, \m{b}) + \m{b} \right] ,
\end{equation*}
where $\m{S}(\beta^{*}) = \left[\left(1 + \sum_{m=1}^M \beta^{*}_m\right)\m{I} - \sum_{m=1}^M \beta^{*}_m \m{W}_{m} \right]^{-1}$ is the matrix of marginal effects and $\mu_{i}(\m{x}, \m{b})$ depends only on $\m{x}$ and $b_{i}$. \cite{Blume_2015_JPE} arguments that the unobserved component $\mu_{i}(\m{x}, \m{b})$ provides the theoretical foundation for the endogeneity caused by the network formation process. As we show below, this is due to individuals choosing their connections through matching via observed and unobserved characteristics.

\vspace{0.3cm}

\noindent \textbf{First Stage.} Individual $i$ chooses to connect with $j$ following a sequential multilayer network formation process. To simplify the process, we assume that individuals are myopic and only consider their connections in layer $m-1$ when forming a relationship in $m$. In our empirical application, this means that the same-grade cohorts represent the network $G_1$ and are formed exogenously first. The adolescents then choose with whom to be friends in the friendship network $G_2$, based on the connections from the classmates.

Individuals $i$ and $j$ choose to connect on layer $m$ to maximize
\begin{equation*}
    \begin{aligned}
    U_{i,m}\left(\m{W}_{m-1}\right) = \sum_{j=1}^n & \ \alpha_{x} | x_{i} - x_{j} | + \alpha_{b} | b_i - \mathbb{E}[b_j | x_i, x_j, b_i] | \\
    & + \alpha_c \m{1}\{i \in C_k\} \m{1}\{j \in C_l\} \sum_{j} w_{ij, m-1} + v_{ij,1} \ , 
    \end{aligned}
\end{equation*}
where $\alpha_x$ represents a homophily parameter over the public characteristics $x$, $\alpha_b$ for the private characteristics $b$, and $\alpha_c$ describes the tendency of individuals to bridge structural holes over the disjoint clusters $C_k$ and $C_l$. In our empirical application, this denotes the utility of creating friendships with a student who is from a different grade-cohort $C_l$. The more classmates a individual $j$ has, the more utility individual $i$ earns if becoming friends. This simplified two-stage social interaction game provides the microfoundation for social conformity in the structural equation \eqref{eq:structural}.


\section{Identification and Estimation Details}\label{app:details}

\subsection{Identification}

The following assumptions are based on \cite{Estrada_2021_WP} adapted to the identification results in \cite{Kuersteiner_2020_ECMA}.

\begin{assumption}[Invertibility]
\label{ass:invertibility}
Define the matrix $\m{S}(\beta^{*}) = \left[\left(1 + \sum_{m=1}^M \beta_m^{*}\right)\m{I} - \sum_{m=1}^M \beta_m^{*} \m{W}_{m} \right]^{-1}$ and assume that $\m{S}(\beta^{*})$ exist for $\beta^{*} \in \Theta_{\beta^{*}}$, where $\Theta_{\beta^{*}} = \Theta_{\beta^{*}_1} \times \dots \times \Theta_{\beta^{*}_M}$ is a compact set.
\end{assumption}

The assumption \ref{ass:invertibility} is used in Eq. \eqref{eq:structural} to put a restriction on the parameter space of $\beta^{*}$ and helps identify structural parameters $\theta$. Under this assumption, we can recover the structural parameters from the estimating equation as 
\begin{equation*}
\beta_m^{*} = \frac{\beta_m}{1 - \beta_f - \beta_c} \ , \quad \delta_m^{*} = \frac{\delta_m}{1 - \beta_f - \beta_c} \ , \quad \text{and} \quad \gamma^{*} = \frac{\gamma}{1 - \beta_f - \beta_c}
\end{equation*}
for $m=f,c$.

For completeness, we provide a formal definition of our identification strategy using the results of \cite{Estrada_2021_WP} and \cite{Kuersteiner_2020_ECMA}.

\begin{assumption}[Moment Restriction]
\label{ass:moment}
$\mathbb{E}\left[\m{x}_{i} e_{i}\right] = 0$ for all $i=1,\dots,n$ and $s \leq t$.
\end{assumption}

Assumption \ref{ass:moment} put a restriction between errors and covariates only for the same individual, in contrast to the classical assumption of $\mathbb{E}\left[\m{x}\ e\right] = 0$. Between individuals $i$ and $j$ we will impose a weak neighborhood dependence assumption instead.

With the multilayer measure of distance $d^{\mathcal{M}}(i,j)$, we define the collection of pairs $\mathcal{P}(a,b,s) = \{ (A, B) : A, B \in \mathcal{N}_n, |A|=a, |B|=b, d_{n,t}(A,B) \geq d \}$, where $d_{n,t}(A,B) = \min \{ d^{\mathcal{M}}(i,j): i \in A, j \in B \}$

To establish \cite{Kojevnikov_2021_JoE} definition of weak dependence, define the $(k \times 1)$ vector $\m{r}_{i} = [\m{x}_{i}^{\top}, e_{i}] \in \mathbb{R}^{k+1}$, and $\m{r}_{A} = (\m{r}_{i} : i \in A)$. 
Let $\mathcal{L}_d,a$ denote the set of bounded real Lipschitz functions mapping $\mathbb{R}^{d \times a} \rightarrow \mathbb{R}$.

\begin{definition}[$\psi-$dependence]
\label{def:psi}
A triangular array $\{\m{r}_{i}\}_{i=1}^n$ is $\psi$-dependent if there exists (1) a sequence $\{\phi_{n,s}\}_{s,n \in \mathbb{N}}$ with $\phi_{n,0}=1$ such that $\sup_n \phi_{n,s} \rightarrow 0$ as $s \rightarrow \infty$ and $\exists S \leq n $ such that if $s>S$ then $\phi_{n,s} = 0$; (2) collection of functionals $\{\psi_{a,b}\}_{a,b \in \mathbb{N}}$ with $\psi_{a,b} : \mathcal{L}_{v,a} \times \mathcal{L}_{v,b} \rightarrow [0,\infty)$ such that
\begin{equation*}
    |\operatorname{Cov}\left(f(\m{r}_{A,t}), g(\m{r}_{B,t})\right)| \leq \psi_{a,b}(f,g) \ \phi_{n,s}
\end{equation*}
for all $A, B \in \mathcal{P}_{n,t}(a,b,s)$, and $f \in \mathcal{L}_{v,a}, g \in \mathcal{L}_{v,b}$.
\end{definition}

Definition \ref{def:psi} uses the sequence $\phi_n$ as the dependence coefficients of $\{\m{r}_{i}\}$. It also states that when two set of nodes are apart from at least a distance $S$ then they are independent. With this definition of $\psi-$dependence we state our weak dependence assumption.

\begin{assumption}[Weak Dependence]
\label{ass:dependence}
For all networks $\mathcal{M}$ that occur with positive probability in $\mathcal{F}$, the conditional distribution $\mathcal{F}(\mathcal{M}, \m{X}, \m{e})$ is such that
\begin{enumerate}
    \item[(i)] $\{\m{r}_{i}\}$ is $\psi-$dependent with dependent coefficients $\phi_n$
    \item[(ii)] For some constant $C>0$, $\ \psi_{a, b}(f, g) \leq C \times a b\left(\|f\|_{\infty}+\operatorname{Lip}(f)\right)\left(\|g\|_{\infty}+\operatorname{Lip}(g)\right)$
\end{enumerate}
\end{assumption}

Assumption \ref{ass:relevance} states necessary conditions for identification. Define $\eta_{m,i}$ and $\eta_{m,i,\lambda}$ as an indicator of whether individual $i$ is non-isolated in the network $\m{W}_{m}$ and $\mathcal{W}_{m, \lambda}$ respectively. In addition, let $\m{D}$ be the matrix with the variables of the rhs of equation \ref{eq:regression}.

\begin{assumption}[Relevance]
\label{ass:relevance} 
Suppose
\begin{enumerate}
    \item[(i)] the event $\eta_{m,i}=0$ and $\eta_{m,i,\lambda}=0$ for all $m,i,\lambda$, happens with probability zero.
    \item[(ii)] $\m{Q}_{ZD} = \operatorname{plim} \frac{1}{n} \m{Z}^{\top} \m{D} < \infty$ and $\m{Q}_{ZX} = \operatorname{plim} \frac{1}{n} \m{Z}^{\top} \m{X} < \infty$
\end{enumerate}
\end{assumption}

\noindent Before showing the theorem for identification, define $\m{S} = n^{-1}\left[\m{y}^{\top} \m{M}_z^{\top} \m{A}_r \m{M}_z \m{y}, \ \m{Wy}^{\top} \m{M}_z^{\top} \m{A}_r \m{M}_z \m{Wy}\right]
$ and $\m{M}_z=\m{I}-\m{D}\left(\m{D}^{\top} \m{P}_z \m{D}\right)^{-1} \m{D}^{\top} \m{P}_z$ with $\m{P}_z=\m{Z}\left(\m{Z}^{\top} \m{Z}\right)^{-1} \m{Z}^{\top}$.

\begin{theorem}[Identification]
\label{the:identification}
Let Assumptions \ref{ass:invertibility}, \ref{ass:moment}, \ref{ass:dependence}, and \ref{ass:relevance} hold for some $K_c$ and $K_d$ such that $K_d \geq K_c + 1$. The parameters $\theta^0 = [\beta^0, \delta^0, \gamma^0]^{\top}$ are identifiable if
\begin{enumerate}
    \item[(i)] $\m{Q}_{ZD}$ has full column rank then $\operatorname{plim} \frac{1}{\sqrt{n}} \m{m}_{l}(\theta) = 0$ (linear moment conditions) has a unique solution at $\theta = \theta^0$.
    \item[(ii)] Only $\m{Q}_{ZX}$ and $\m{S}$ has full column rank then $\operatorname{plim} \frac{1}{\sqrt{n}} \m{m}(\theta) = 0$ (linear and quadratic moment conditions) has a unique solution at $\theta = \theta^0$.
\end{enumerate}
\end{theorem}

\noindent The proof for Theorem \ref{the:identification} follows from \cite{Kuersteiner_2020_ECMA} under their Lemma EX1.


\subsection{Average Social Effects}

In this section, we detail the process to obtain standard errors for the individual social effects. Recall that the matrix of social effects is defined as $\m{S}(\beta^{*}) = \left[\left(1 + \sum_{m=1}^M \beta_m^{*}\right)\m{I} - \sum_{m=1}^M \beta_m^{*} \m{W}_{m} \right]^{-1}$. However, we want to rewrite the matrix of social effects using the estimated coefficients as $\m{S}(\hat{\beta}) = (1 - \hat{\beta}_1 - \hat{\beta}_2)(\m{I} - \hat{\beta}_1 \m{W}_1 - \hat{\beta}_2 \m{W}_2)^{-1}$. We aim to obtain standard errors for the vector of average social effects $\bar{s}_j = n^{-1} \sum_{j \neq i} \hat{s}_{ij}$. To apply the delta method we need $\partial \bar{s}_j / \partial \beta$, which involves to take the derivate of the inverse of a sum of matrices. To overcome that, we express the matrix of social effects in terms of the infinite sum of the product of the different adjacency matrices, given by 
\begin{equation*}
    \m{S}(\hat{\beta}) = (1 - \hat{\beta}_1 - \hat{\beta}_2) \sum^{\infty}_{r=0} \left(\hat{\beta}_1 \m{W}_1 + \hat{\beta}_2 \m{W}_2\right)^{r} .
\end{equation*}

Therefore, we can calculate the variance-covariance matrix for the average social effects as
\begin{equation*}
    \m{V(\bar{s})} = \frac{\partial \bar{s}}{\partial \beta}^{\top} \widehat{\m{\Sigma}}_{\beta} \frac{\partial \bar{s}}{\partial \beta}
\end{equation*}
where 
\begin{equation*}
    \frac{\partial}{\partial \beta_m} \m{S}(\hat{\beta})=-\sum_{r=0}^{\infty}\left(\hat{\beta}_1 \m{W}_1 + \hat{\beta}_2 \m{W}_2\right)^r+\left(1-\beta_1-\beta_2\right)\left[\sum_{r=1}^{\infty} r \m{W}_m\left(\hat{\beta}_1 \m{W}_1 + \hat{\beta}_2 \m{W}_2\right)^{r-1}\right]
\end{equation*}

\end{document}